\documentclass[prd,onecolumn,showpacs,nofootinbib,preprintnumbers,superscriptaddress]{revtex4}

\usepackage{amsmath}
\usepackage{amsfonts}
\usepackage{graphicx}
\usepackage{hyperref}
\usepackage{mathbbol}
\usepackage{lipsum}
\usepackage{xcolor}
\usepackage{tabularx}
\usepackage{multirow}
\usepackage{makecell}
\usepackage{cancel}
\usepackage{booktabs}
\allowdisplaybreaks

\def\be{\begin{equation}}   
\def\ee{\end{equation}}  
 \def\bea{\begin{eqnarray}}    \def\eea{\end{eqnarray}}  \def\no{\nonumber}    
    \def\d{{\rm d}}   \def\m{{\rm m}}             
\def\r{\right}            \def\l{\left}

\begin{document}
\title{Cosmological dynamics of holographic dark energy with non-minimally coupled scalar field}
\date{\today}
\author{Amornthep Tita}\email{amornthepti61@nu.ac.th}
 \affiliation{NAS, Centre for Theoretical Physics \& Natural Philosophy, Mahidol University, Nakhonsawan Campus,  Phayuha Khiri, Nakhonsawan 60130, Thailand}
  \affiliation{The Institute for Fundamental Study ``The Tah Poe Academia Institute", Naresuan University, Phitsanulok 65000, Thailand}
 \author{Burin Gumjudpai}\email{Corresponding: burin.gum@mahidol.ac.th}
 \affiliation{NAS, Centre for Theoretical Physics \& Natural Philosophy, Mahidol University, Nakhonsawan Campus,  Phayuha Khiri, Nakhonsawan 60130, Thailand}
\author{Pornrad Srisawad}\email{pornrads@nu.ac.th}
\affiliation{Department of Physics, Naresuan University, Phitsanulok 65000, Thailand}

\begin{abstract}
In this study, we consider FRW universe filled with matter, non-minimally coupling (NMC) scalar field under $V(\phi) = V_{0}\phi^{2}$ potential and holographic vacuum energy. Dark energy is contributed from both holographic vacuum energy and the NMC scalar field. NMC effective gravitational constant $G_\text{eff}(\phi)$, is naturally defined at the action level. Therefore, the gravitational constant in the holographic vacuum density is an effective one, i.e.
$
 \rho_{\Lambda} = {3c^{2}}/{8\pi G_{\text{eff}}L^{2}}\,.
$
Apparent horizon is chosen as IR holographic cutoff scale as it is a trapped null surface.   There are nine fixed points in this dynamical system with four independent dimensionless parameters. 
 We consider flat case and find that viable cosmological evolution follows the sequence: an initial stiff-fluid-dominated phase, transitioning through a nearly dust-dominated era, and eventually reaching a stable dark energy-dominating state. Stability analysis requires that $\xi <0$ and $0 < c < 1$ for the theory to be physically valid. 
  Since zero NMC coupling, $\xi=0$, is not allowed in the autonomous system, the model can not completely recover canonical scalar field case. 
  That is to say, as $\xi \rightarrow 0^-$ and $c \rightarrow 0^+$, the model can only approach the canonical scalar case but can not completely recover it.
 To approach dust or stiff fluid dominations, both magnitudes of the NMC coupling and the holographic parameter must be small. Numerical integration shows that for any allowed values of $\xi$ and $c$, $w_\text{eff}$ approaches $-1$ at late times. Increasing of $c$ does not change shape of the $w_{\rm eff}$, but larger $c$ increases $w_\text{eff}$.  As $\xi$ becomes stronger, dust era gradually disappears.  Good behaviors of the dynamics require $-1 \ll \xi <0$ and $0 < c \ll 1$.

\end{abstract}

\date{\today}

\vskip 1pc

\maketitle \vskip 1pc

\section{Introduction}  \label{intro}

Over recent decades, numerous theoretical frameworks have emerged to address cosmological key puzzles such as the late-time universe acceleration with $w\simeq -1$ \cite{Amanullah2010, Astier:2005qq, Goldhaber:2001a, Perlmutter:1997zf, Perlmutter:1999a, Riess:1998cb, Scranton:2003, Tegmark:2004}, graceful exits and origin of structure from inflationary scenarios which is consistent with the observed Cosmic Microwave Background (CMB) data  \cite{WMAP9, Aghanim:2019ame}. Two predominant strategies to tackle these puzzles is to modify either the matter sector or the gravitational sector \cite{VF,CL,CF,odin,en,Ishak:2018his,MaedaBook}. One of the idea is a mixing modification of matter and gravitation. This is to allow 
non-minimal coupling between scalar field to Ricci scalar in form  of $\sqrt{-g}f(\phi)R$ as in Jordan--Brans--Dicke models \cite{Brans:1961sx, Dirac}. 
 Non-minimal coupling term is also motivated by re-normalizing term of quantum field in curved space \cite{Davies} or supersymmetries, superstrings and induced gravity theories \cite{Zee, Cho, Salam, Accetta, Maeda}. It has been considered 
 in extended inflation with first-order phase transition and other inflationary models
 \cite{ei, fs, y, ac, fm, kasper, Amendola:1993it} with possibility in describing of the late acceleration \cite{q, Uzan:1999ch, e, Easson:2006jd, Amendola:1999qq, CapozzRitis}.

Considering $\frac{1}{2}\xi \phi^{2}R$ term, 
this is known as non-minimally coupling theory (NMC), where $\xi$ is coupling constant. The NMC
 is realized as a subclass of generalized scalar-tensor theory, the Horndeski theory  with the action \cite{Horndeski:1974wa,Kobayashi:2011nu},
\begin{eqnarray}
S= \int \d^{4}x \sqrt{-g} \Biggl[\sum_{i=2}^{5} \frac{1}{8\pi G} \mathcal{L}_{i} + \mathcal{L}_{\rm m}\Biggr]\,.
\end{eqnarray}
For the NMC theory, $ 
\mathcal{L}_{2}  =  
\mathcal{L}_{3}  =  \mathcal{L}_{5}  = 0 $ and the forth Horndeski Lagrangian, $
\mathcal{L}_{4} \, = \,  G_{4}(\phi,X)R + G_{4,X}[(\Box\phi^{2}) - (\nabla_{\nu}\nabla_{\mu}\phi) (\nabla^{\nu}\nabla^{\mu}\phi) ] $
takes the form,  $
G_{4} = ({1}/{2})\big[ {(8\pi G)}^{-1} - \xi \phi^{2} \big] \ 
$ with $ G_{4,X} = 0 $. In this Lagrangian, $X \equiv \phi^{\mu}\phi_{\mu}$ and $\Box \equiv \nabla_{\mu}\nabla^{\mu}$.  Difference between NMC and other simplest scalar-tensor theory such as Jordan--Brans--Dicke theory, is that the NMC contains non-minimal form of $G_{4}$. On the other hand, the Jordan--Brans--Dicke theory has minimal form $G_{4} = f(\phi)/16\pi G$ added with $G_{2} = (\omega/\phi)\phi^{,\mu}\phi_{,\mu}$  \cite{Brans:1961sx}.
The NMC action in the Jordan frame with matter and vacuum energy is 
\begin{equation}
S = \int \d^{4}x \sqrt{-g} \left[\frac{R}{16\pi G} - \frac{1}{2}\xi \phi^{2}R - \frac{1}{2}g^{\mu\nu}\nabla_\mu\phi\nabla_\nu\phi + \mathcal{L}_{\rm m} + V(\phi) \right]   +   S_{\Lambda}\,. \label{NMCaction}
\end{equation}
In this paper we use metric signature of $(-+++)$, and $\hbar = c = 1$.
At Lagrangian level, one can view the theory with effective gravitational constant in the Einstein field equation as 
\begin{eqnarray}
G_{\rm eff} = \frac{G}{1- 8\pi G\xi \phi^{2}}\,. \label{GeffNMC}
\end{eqnarray}
Benefits for having $G_{\rm eff}$ can be seen in many aspects, first of all it is possible to have a new locally-scale invariant extension of general relativity based on Weyl geometry \cite{Smolin:1979uz}. Secondly, instabilities induced by NMC coupling can cause cosmological constant damping so that it matches present-day value of cosmological constant \cite{Ford:1987de}.
Moreover, there are many studies devoted to obtain viable cosmological models from the NMC theory \cite{deRitis:1989zm,Uzan:1999ch,Perrotta:1999am,Gupta:2009kk,Sami:2012uh,Yi:2016jqr}.

Several inflationary models are considered to constrain the NMC coupling parameter $\xi$, such as chaotic inflation \cite{fm,Fakir:1990eg,Makino:1991sg}, Higgs inflation \cite{Bezrukov:2007ep,Hertzberg:2010dc}, generic potential $ V(\phi) = \alpha \phi + \frac{1}{2}m\phi^{2} + \beta \phi^{3}  + \frac{1}{4}\lambda\phi^{4} - \Lambda$ \cite{Bertolami:1987wj}, monomial potential \cite{Hrycyna:2020jmw}, inverse potential \cite{Shahalam:2020lcc} and quadratic potential \cite{Eshaghi:2015rta}.
NMC inflationary parameters in Einstein frame are constrained with WMAP5+SDSS+SNIa data \cite{Nozari:2010uu}\footnote{sign of the coupling $\xi$ in \cite{Nozari:2010uu}
is defined opposite from ours.}. In both Einstein and Jordan frames, it is found that up to the second order of the slow-roll parameters expressed in spectral index, the coupling is in a range, $\xi \leq -0.0003$ and 
$ -0.1666  \leq \xi $  for $V(\phi) \propto \phi^{p}$ with $p = 4$ \cite{Nozari:2010uu}
(see also in \cite{Nozari:2007eq}). 
It is found also that for $p=4$
viable cosmological models exist even for $  \xi \ll -1$ 
\cite{Bez2008, Bez2009, Simon2009, bez2009(2), Bar}. When considering cosmological perturbation of the NMC model, for $p=2$, the constraint is $\xi > -7.0 \times 10^{-3}$ and for  $p=4$,  the model is disfavored for $\xi = 0$, i.e. minimal coupling case. Having the NMC term can rescue the $p=4$ case but with a observational constraint of  $\xi < -1.7 \times 10^{-3}$ \cite{Tsujikawa:2004my}. In our work, we consider the case of $p=2$ with $\xi > -7.0 \times 10^{-3}$ constraint.

Apart from the problems in cosmology, quest in unifying gravitational theory to quantum theory invites us to the holographic principle which is a possible way in explaining small observed value of present-day vacuum energy.  Dark energy models incorporating the holographic principle applied to late cosmic acceleration are known as Holographic Dark Energy (HDE) models. 
Holographic principle, as postulated by t'Hooft \cite{tHooft:1993dmi}, suggests that the information about a volume of space could be encoded on its corresponding boundary, just like a hologram. Following 't Hooft's idea, Susskind emphasized that in context of string theory,  entropy of a volume is proportional to the area of its surface boundary \cite{Susskind:1994vu}.
Considering a black hole, holographic principle becomes a key to understand how information of a three-dimensional black hole is encoded on its surface of  two-dimensional event horizon. 
Black hole information and its event horizon are related by the Bekenstein--Hawking entropy, $S_{\text{BH}} = \pi L^2 M_{\text{P}}^2 $ where $L$ is the Schwarzschild radius and $M_{\rm P} = (8\pi G)^{-1/2}$ is the reduced Planck mass
\cite{Hawking:1971tu,Bekenstein:1973ur,Bekenstein:1974ax,Hawking:1974rv,Hawking:1975vcx}. A matter volume with entropy exceeding this limit turns into a black hole.  
Considering field theory in a box sized $L$ with UV cutoff $\Lambda$, field-theoretic derived entropy of the volume scales extensively as $S \sim L^{3}\Lambda^{3}$. 
The entropy derived with field theoretic method is hence bounded by the Bekenstein limit, $L^{3}\Lambda^{3} \lesssim S_{BH}  = \pi L^{2}M^{2}_{\rm P}$. 
t'Hooft  and Susskind
 found that field-theoretic derived entropy is larger than it should due to some over-counting degrees of freedom \cite{tHooft:1993dmi,Susskind:1994vu} 
In 1999 Cohen, Kaplan, and Nelson proposed a more restrictive limit than the Bekenstein limit. This is, $L^{3}\Lambda^{4} \lesssim \pi L M_{\rm P}^{2} \rightarrow \rho_{\Lambda} \left(4/3\right)\pi L^{3} = L/(2G)$ which is later known as Cohen--Kaplan--Nelson (CKN) bound \cite{Cohen:1998zx}.
Let $\rho_{\Lambda}$ denotes zero-point quantum energy density, sum of zero-point one-loop correction of energy density in effective field theory is saturated by the CKN bound as  
\begin{eqnarray}
\rho_{\Lambda} = \frac{3c^2}{8\pi G L^2} \ , \label{HDEpure}
\end{eqnarray}
where $L$ is infrared cutoff scale.
One could suggest that infrared cutoff scale could be as large as the Hubble length $L = H^{-1}$, resulting that $\rho_{\Lambda}$ is comparable to the dark energy in the present day. Therefore it is considered as dark energy \cite{Hsu:2004ri}. However this setting leads to a dust-like equation of state, hence other choices of the infrared cutoff are considered to get away from this problem. Many alternative cutoffs were proposed such as future event horizon cutoff \cite{Li:2004rb}, conformal time cutoff (dubbed agegraphic dark energy model) \cite{Cai:2007us}, Ricci scalar of FLRW spacetime (dubbed Ricci dark energy model) \cite{Gao:2007ep}, Granda-Oliveros cutoff \cite{Granda:2008tm} and many others.

Early work on application of the holographic dark energy to NMC model, in Jordan frame, was considered by Ito \cite{Ito:2004qi} using Hubble horizon cutoff with $c=1$ without any free scalar potential. By assuming power-law scale factor and power-law scalar solutions, a viable range of $\xi$ is computed.
In Jordan frame, as NMC theory naturally has effective varying $G$. 
Hence holographic cosmological dynamics is affected by the NMC effective varying gravitational constant. This opens up some possibility of addressing cosmological challenges \cite{Jamil:2009sq}. This is accompanied by the work of Setare \& Saridakis \cite{Setare:2008pc} in NMC-HDE context without $V(\phi)$ of which the consideration includes quintom and phantom extension. 
Similar scenario is considered by Granda and Escobar \cite{Granda:2009zx} using Granda-Oliveros cutoff. Applications of HDE  
with $G_{\text{eff}}$ in NMC sister theory, the 
non-minimal derivative coupling theory (NMDC), are recently reported in \cite{Kritpetch:2020vea,Baisri:2022ivv,Tita:2024jzw}.

This work is organized as follow. In section \ref{NMC-eq}, NMC field equation and holographic vacuum energy density are introduced. In section \ref{dynnnnn},  dynamical variables are defined 
and dynamical analysis is performed. Phase portraits are shown in section \ref{ppt}. Numerical integrating solutions are described in section \ref{num}. Finally conclusion is made in section \ref{conclusion}

\section{NMC Field Equations and Holographic dark energy}  \label{NMC-eq}
Varying the Jordan frame NMC action \eqref{NMCaction} with respect to metric tensor $g^{\mu\nu}$, we obtain the Einstein field equations \cite{Uzan:1999ch},
\begin{eqnarray}
    \left(1-8\pi G\xi\phi^2\right)G_{
\mu\nu}=8\pi G \left[			\nabla_\mu\phi\nabla_\nu\phi
			-\frac{1}{2}g_{\mu\nu}\nabla^\rho\phi\nabla_\rho\phi
			+V(\phi) g_{\mu\nu}-\xi\left(\nabla_\mu\nabla_\nu\phi^2-g_{\mu\nu}\nabla_\rho\nabla^\rho\phi^2\right)
			+T^{(\text{m})}_{\mu\nu}
            +T^{(\Lambda)}_{\mu\nu}
        \right]\,.
        \label{NMC_EFE}
\end{eqnarray}
Matter content in this work includes the NMC scalar field, non-relativistic matter (or dust), and holographic vacuum energy.  
By varying the action \eqref{NMCaction} with respect to the scalar field $\phi$, we obtain the Klein-Gordon equation for the NMC action as
    \begin{equation}
        \nabla_\mu\nabla_\nu\phi-\xi R\phi=0\,.
        \label{NMC_KG}
    \end{equation}
In a spatially curved FLRW universe, $(0,0)$ and $(i,j)$ components of the field equations gives the Friedmann equations as follows:
\begin{equation}
    3H^{2} + \frac{3k}{a^{2}}  =  8\pi G_\text{eff} \left( \frac{\dot{\phi}^{2}}{2} + V(\phi) + 6\xi H \phi \dot{\phi} + \rho_{\rm m} + \rho_{\Lambda} \right) \,, \label{FriedmannNMC}
\end{equation}
\begin{widetext}
\begin{equation}
    2\dot{H}+3H^{2} + \frac{k}{a^{2}} = -8\pi G_{\text{eff}}\left[\frac{\dot{\phi}^{2}}{2}-V(\phi)  -2\xi\left(\dot{\phi}^{2} +\phi\ddot{\phi} + 2\phi H\dot{\phi}\right)  + P_{\rm m} + P_{\Lambda} \right]\,. \label{accelNMC}
\end{equation}
\end{widetext}
Moreover, the Klein-Gordon equation is given by
\begin{equation}
\ddot{\phi} + 3H\dot{\phi} = -V_{,\phi} + 6\xi \left( \dot{H} + 2H^2 + \frac{k}{a^2}
 \right)\phi \,, \label{KleinGordonNMC}
\end{equation} 
where $V_{,\phi}$ denotes the derivative with respect to $\phi$. In NMC theory, the gravitational constant becomes a function of the NMC scalar field, expressed as $G_\text{eff}(\phi)$. Consequently, the holographic vacuum energy density is given by 
 \begin{eqnarray}
 \rho_{\Lambda} = \frac{3c^{2}}{8\pi G_{\text{eff}}L^{2}}\,,
 \end{eqnarray}
 where $L$ is the holographic cutoff and $0 \leq c<1$. In the holographic scenario, a cosmological bulk region should be enclosed by a trapped null surface, similar in concept to a black hole event horizon, 
where light never reaches the horizon. The apparent horizon, which is a trapped null surface in an accelerating expansion, is a natural choice for the holographic cutoff. It has been shown that the first law of thermodynamics at the apparent horizon can connect to the Friedmann equation \cite{CaiKim}.
The cutoff is hence taken to be the apparent horizon: 
 \begin{eqnarray}
 L = \frac{1}{\sqrt{H^{2} + \dfrac{k}{a^{2}}}}\,.
 \end{eqnarray}
This gives the holographic vacuum density,
\begin{equation}
\rho_{\Lambda} = \frac{3c^{2} \left(1-8\pi G \xi\phi^{2} \right)\left(H^{2} + \dfrac{k}{a^{2}} \right)}{8\pi G} \,.  \label{NMC_HDE}
\end{equation} 
Next section, we will further investigate its dynamical behavior under power-law potential.


\section{Dynamics of the Holographic NMC model}  \label{dynnnnn}
Considering power-law potential given by $V(\phi) = V_{0}\phi^{n}$ in performing dynamical analysis, we define dimensionless dynamical variables as follows, 
\begin{equation}
	\begin{split}
	&
	x\equiv\frac{8\pi G_\text{eff}\dot{\phi}^2}{6H^2},\;\quad 
	y\equiv\frac{8\pi G_\text{eff} V_0 \phi^n}{3H^2},\;\quad
	s\equiv\frac{16\pi G_\text{eff}\xi\phi\dot{\phi}}{H},\;\quad
	\Omega_{\rm m}\equiv\frac{8\pi G_\text{eff}\rho_{\rm m}}{3H^2},\;\quad 
    \\&	
    \Omega_\Lambda\equiv\frac{8\pi G_\text{eff}\rho_\Lambda}{3H^2},\;\quad\;
	\Omega_k\equiv-\frac{k}{a^2H^2}\;\;\;\text{and}\;\;\; A\equiv8\pi G_\text{eff}\xi\phi^2\,,
	\end{split}
\label{dim less para}
\end{equation}
where $\Omega_\text{m}, \Omega_\Lambda$, and $\Omega_k$ are the density parameters for the dust matter, holographic vacuum energy, and spatial curvature, respectively. An additional parameter, $A$, is introduced to make the dynamical system autonomous. 
The parameter $x$ represents the scalar field kinetic term, $y$ denotes the power-law potential term. 
Effectively, NMC effect is within $G_\text{eff}$ in almost all dynamical variables via $G_\text{eff}$. 
The variable $s$ represents the NMC coupling term. Notably, $s=0$ implies $\xi=0$ and this changes $G_\text{eff}$ back to $G$ in all dynamical variables, recovering non-NMC limit of the model. On the other hand $c=0$ implies no holographic vacuum energy.
The Friedmann equation \eqref{FriedmannNMC} can be written in terms of dynamical parameters and expressed as a constraint equation, normalizing the density parameters of all density components,  
\begin{eqnarray}
    1&=&\; \Omega_\phi+\Omega_{\rm m}+\Omega_\Lambda+\Omega_k\,.
    \label{FM_constraint}
\end{eqnarray}
The density parameter of the scalar field, which is non-minimally coupled to the Ricci scalar, can be expressed as
    \be
        \Omega_\phi=x+y+s\,.
        \label{scal_den}
    \ee
From equation \eqref{FM_constraint}, both the holographic vacuum energy and the scalar field drive the acceleration such that dark energy density parameter is 
    \be
        \Omega_\text{DE}  \equiv \Omega_\phi+\Omega_\Lambda\,.
    \ee
According to the energy density of HDE in equation \eqref{NMC_HDE}, there is a relationship between the HDE density parameter and the spatial curvature term. This results in the constraint
    \begin{equation} 
	\Omega_\Lambda=c^2(1-\Omega_k)\,.
	\label{holographic_constraint}
    \end{equation}
This relation is independent of the NMC coupling. Additionally, there is another constraint arising from the relationship between $x$, $s$, and $A$,
\begin{equation}
    x=\frac{s^2}{24\xi A}\,,
    \label{x,s,A_constraint}
\end{equation}
where $\xi \neq 0$, otherwise indeterminate.
Hence, the Friedmann constraint provides an expression for the dust density parameter in terms of $y,\; s,\; A$ and $\Omega_k$, as 
\begin{equation}
	\Omega_{\rm m}=1-\frac{s^2}{24\xi A}-y-s-c^2(1-\Omega_k)-\Omega_k\,.
	\label{Friedmann_constraint_2}
\end{equation}
Equation of state parameter for a NMC scalar field is given by using its energy density and pressure. According to equations \eqref{FriedmannNMC} and \eqref{accelNMC}, this gives
\begin{align}
    \rho_\phi&=\frac{\dot{\phi}^{2}}{2} + V(\phi) + 6\xi H \phi \dot{\phi}\,,
    \\
    P_\phi&= \frac{\dot{\phi}^{2}}{2}-V(\phi)  -2\xi\left(\dot{\phi}^{2} +\phi\ddot{\phi} + 2\phi H\dot{\phi}\right)\,,
    \\
    w_\phi&=\frac{P_\phi}{\rho_\phi}=\frac{s+3[x-4x\xi+y(2n\xi-1)-4A\xi(\epsilon+\Omega_k-2)]}{3(x+y+s)}\,.
    \label{scalar_field_eos}
\end{align}
Using equation \eqref{NMC_HDE} together with the continuity equation of holographic vacuum energy, $\dot{\rho}_\Lambda+3H\rho_\Lambda(1+w_\Lambda)=0$, equation of state parameter of holographic vacuum energy equation is  
    \be
        w_\Lambda=\frac{-1+2\epsilon/3+\Omega_k/3+s(1-\Omega_k)/3}{1-\Omega_k}\,.
        \label{HDE_eos}
    \ee
where $\epsilon\equiv -\dot{H}/H^2$. Equation of state parameter for dark energy is defined here,   
    \be
        w_\text{DE}=\frac{w_\phi\Omega_\phi+w_\Lambda\Omega_\Lambda}{\Omega_\phi+\Omega_\Lambda}.
    \ee
According to equations \eqref{dim less para}, \eqref{holographic_constraint}, \eqref{x,s,A_constraint}, and \eqref{Friedmann_constraint_2}, we have seven dynamical variables and three constraint equations. Consequently, we can construct the dynamical system using only four independent variables: $y,\; s,\; A,$ and $\Omega_k$.

\subsection{Autonomous system}
The autonomous system is constructed by taking the derivative of the dynamical parameters with respect to e-folding number, $N \equiv \ln a$. This is  
    \begin{align}
  \no   y'&=sy+\frac{nsy}{2A}+2y\epsilon\,,\\
  \no   s'&=s^2+\frac{s^2}{2A}+s(\epsilon-3)-6ny\xi+12A\xi(\epsilon-2+\Omega_k)\,,\\
  \no  A'&=s(1+A)\,,\\
     \Omega_k'&=-2\Omega_k(1-\epsilon)\,.    
    \label{eq: autonomous system}
    \end{align}
Here, $``\; '\; "$ denotes derivative with respect to $N$. The parameter $\epsilon\equiv -\dot{H}/H^2$ is expressed in terms of the dynamical variables to obtain the autonomous system. Using equation \eqref{accelNMC} together with the NMC Klein-Gordon equation \eqref{KleinGordonNMC} and equation of state parameter of holographic vacuum energy equation \eqref{HDE_eos},   
\begin{equation}
    \epsilon=\frac{s^2\left(1-4\xi\right)+8As\xi\left[1+c^2(1-\Omega_k)\right]-8A\xi\left[-3+3y-24A\xi-6ny\xi+c^2(3-\Omega_k)+\Omega_k+12A\xi\Omega_k\right]}{16A\xi(1-c^2+6A\xi)}.
\end{equation}
Using the Friedmann equation \eqref{FriedmannNMC}, expressed in the form $3H^2+3k/a^2=8\pi G_\text{eff}\rho_\text{tot}$, and equation \eqref{accelNMC}, expressed as $2\dot{H}+3H^2+k/a^2=-8\pi G_\text{eff}P_\text{tot}$, the effective equation of state parameter is hence,
    \begin{equation}
        w_\text{eff} \,=\,  \frac{P_\text{tot}}{\rho_\text{tot}} \,= \,    -\frac{2\dot{H}+3H^2+k/a^2}{3H^2+3k/a^2}=\frac{-1+2\epsilon/3+\Omega_k/3}{1-\Omega_k}\,.
      \label{eff_EOS}
    \end{equation}
Fixed points of the autonomous system are found by setting equation \eqref{eq: autonomous system} to zero. A list of the fixed points for all dynamical variables is shown in TABLE \ref{table_fixedpoint}. To determine stability of each fixed point, we analyze the system using linear perturbations about each fixed point, i.e. $y=y_c+\delta y,\; s=s_c+\delta s,\; A=A_c+\delta A,\; \Omega_k=\Omega_{kc}+\delta\Omega_k$, and examine the eigenvalues of the resulting Jacobian matrix $\mathcal{M}$, 
    \begin{equation}
        \frac{{\rm d}}{{\rm d}N}
        \begin{pmatrix}
            \delta y \\
            \delta s \\
            \delta A \\
            \delta \Omega_k
        \end{pmatrix} = \mathcal{M}
	\begin{pmatrix}
		\delta y\\
		\delta s\\
		\delta A\\
		\delta\Omega_k
	\end{pmatrix}\;, 
\end{equation}
where $y_c, s_c, A_c, \Omega_{kc}$ denote fixed points and $\delta y, \delta s, \delta A, \delta \Omega_k$ represent linear perturbations.
The Jacobian matrix $\mathcal{M}$ is defined as  
\begin{equation}
	\mathcal{M}=\begin{pmatrix}
		
     \partial_y y' & \partial_s y'  &\partial_A y' &\partial_{\Omega_k} y'
		\\
    \partial_y s' & \partial_s s'  &\partial_A s'  &\partial_{\Omega_k} s'
		\\
    \partial_y A' & \partial_s A'  &\partial_A A'  &\partial_{\Omega_k} A'
		\\
    \partial_y \Omega_k' & \partial_s \Omega_k'  &\partial_A \Omega_k'  &\partial_{\Omega_k} \Omega_k'		
\end{pmatrix}_\text{at\;fixed\;points}
\label{Jacobian_matrix}
\end{equation}
where, for example, $\partial_y y', \partial_s y', \partial_A y', \partial_{\Omega_k} y'$ denote partial derivative of $y'$ with respect to $y,s,A,$ and $\Omega_k,$ respectively. The eigenvalues of the Jacobian matrix reveal the stability characteristics of the fixed points. Since the autonomous system has four degrees of freedom, the Jacobian matrix has four eigenvalues.

A fixed point is stable if all eigenvalues are negative, unstable if all eigenvalues are positive, and a saddle point if at least one eigenvalue is positive. When some eigenvalues are zero, linear stability method is not applicable. In such cases, numerical integration is performed using various initial conditions about the fixed point. Late evolution of the dynamical variables then indicates stability of the fixed point.

%
%
\begin{table}
	\renewcommand{\arraystretch}{2.5}
	\begin{tabular}{|c|c|c|c|c|c|c|c|}
		\hline
		\multirow{2}{*}{{Name}}&
            \multicolumn{7}{c|}{{Fixed points}}             \\
		\cline{2-8}
		& $x_c$ &  $y_c$ &  $s_c$  & $A_c$ & $\Omega_{kc}$ & $\Omega_{\Lambda c}$ & $\Omega_{{\rm m}c}$
		\\
		\hline\hline
		1 & 0  & 0 & 0 & $A_c$ &  1 & 0 & 0 
		\\
  \hline
		2 &  $-\dfrac{1}{24\xi}$ & 0 & 1 &  $-1$  & $1-\dfrac{1}{8\xi}$ & $\dfrac{c^2}{8
\xi}$ & $-1+\dfrac{1}{6\xi}-\dfrac{c^2}{8\xi}$   
		\\
  \hline
		3 & $-\dfrac{2B^2}{3c^4\xi}$ & 0 & $\dfrac{4B}{c^2}$ & $-1$ & \makecell{
        $\Big[2(1-6\xi)^2+3c^4\xi$
        \\
        $+2c^2(-1+6\xi)\Big]\Big/ {3c^4\xi}$
        }&  
        $-\dfrac{2(-1+6\xi)B}{3c^2\xi}$ & 0      
		\\
  \hline
		4 & 
        $-\dfrac{3}{8\xi}$  & 
        $2-\dfrac{3}{8\xi}$ & 
        3 & 
        $-1$ & 0  & $c^2$ & $-4-c^2+\dfrac{3}{4\xi}$ 
		\\
  \hline
		5 & 
        $-\dfrac{2\xi}{3(1-4\xi)^2}$ & 0 & $\dfrac{4\xi}{1-4\xi}$ & $-1$ & 0 & $c^2$ & $-c^2+\dfrac{3-34\xi+96\xi^2}{3(1-4\xi)^2}$    
		\\
  \hline
        6 & $-\dfrac{2\xi(1+c^2)^2}{3(1-4\xi)^2}$ & \makecell{$\Big[3-34\xi$\\$+2c^4\xi+96\xi^2$\\$+c^2(-3+16\xi)\Big]\Big/$\\$3(1-4\xi)^2$} & $\dfrac{4\xi(1+c^2)}{1-4\xi}$ & $-1$ & 0 & $c^2$ & 0     
		\\
  \hline
        7 & \makecell{$1-c^2-12\xi$\\$+2\sqrt{6\xi B}$} & 0 & $12\xi-2\sqrt{6\xi B}$ & $-1$ & 0 & $c^2$ & 0 
		\\
  \hline
        8 & \makecell{$1-c^2-12\xi$\\$-2\sqrt{6\xi B}$} & 0 & $12\xi+2\sqrt{6\xi B}$ & $-1$ & 0 & $c^2$ & 0     
		\\
  \hline
        9 & 0 & $1-c^2$ & 0 & $
        \dfrac{-1+c^2}{2}$ & 0 & $c^2$ & 0     
		\\
        \hline
	\end{tabular}
 \caption{Fixed points of all dynamical variables for $n=2$ case, where $B\equiv-1+c^2+6\xi$. Note that for $c=0$ and $\xi = 1/6$, $B=0$.}
    \label{table_fixedpoint}
\end{table}

\begin{table}
    \centering
    \renewcommand{\arraystretch}{3}
    \begin{tabular}{|c|c|c|c|}
    \hline
    \makecell{Fixed\\ points}  & eigenvalues &  stability & \makecell{existence conditions \\ for $0\leq c<1$}\\
    \hline\hline
        1 &  $0$,\;\; $-1$,\;\; $-2$,\;\; $2$ & saddle & $\forall\xi,\; \xi \neq 0$\\
        \hline
       2 &  \makecell{ 
        \vspace{0.2cm}\\ 
        $1$,\;\; $2$,\;\; 
        $\dfrac{-\left[4\xi \left(c^2-4\right) + c^2+96 \xi ^2\right]-\sqrt{ \Delta }}{32 \xi B}$,
        \vspace{0.2cm}\\
        $\dfrac{-\left[4\xi \left(c^2-4\right) +c^2+96 \xi ^2\right]+\sqrt{ \Delta }}{32 \xi B}$ \vspace{0.2cm} } 
    & 
        \makecell{saddle,\vspace{0.75cm}\\ \\ unstable \\ \\} 
    & 
        \makecell{
        if $\xi<0$, or\\ $(4-3c^2)/24<\xi,$ 
        \vspace{0.1cm}
        \\
        for $\xi\neq 0$ and $\xi\neq(1-c^2)/6,$ 
        \vspace{0.3cm}
        \\
        if $0<\xi<(4-3c^2)/24, $
        \vspace{0.1cm}
        \\
        for $\xi\neq 0$ and $\xi\neq(1-c^2)/6$}\\
        \hline
        3 & $2$,\;\; $\dfrac{4B}{c^2}$,\;\; $\dfrac{3 c^2+24 \xi -4}{c^2}$,\;\; $\dfrac{6 c^4 \xi +4 c^2 (6 \xi -1)+4 (1-6 \xi )^2}{3 c^4 \xi}$ &  \makecell{ \\ saddle, \vspace{0.5cm} \\  unstable \\ \\} 
        &
        \makecell{ 
        if $\xi<\big( 8-c^4-4c^2$ 
        \\ $+c^3\sqrt{8+c^2}\big)/48,\;\xi\neq 0, c\neq 0$ 
         \vspace{0.3cm}\\ 
            if $\big( 8-c^4-4c^2$ 
        \\ 
            $+c^3\sqrt{8+c^2}\big)/48\leq\xi,\;\xi\neq 0, c\neq 0$  
        } \\
        \hline
        4 &  \makecell{\\ $-2$,\;\; $3$,\;\; 
        $\dfrac{3 c^2  (4 \xi -1)-\sqrt{\Theta}}{4B}$,\vspace{0.2cm}
          \;\; 
        $\dfrac{3 c^2  (4 \xi -1) + \sqrt{\Theta}}{4B}$ \vspace{0.2cm} } & saddle & $\xi\neq (1-c^2)/6,\;\xi\neq 0$\\
        \hline
        5 & \makecell{$\dfrac{4\xi}{1-4\xi},\;\;
        \dfrac{1-8\xi}{1-4\xi},\;\; 
        \dfrac{3-16\xi}{1-4\xi},\;\;
        \dfrac{3 c^2 (1-4 \xi )^2-96 \xi ^2+34 \xi -3}{2 (4 \xi -1)B }$ \vspace{0.2cm}}  & saddle & \makecell{
        $\xi\neq 1/4\;,\xi\neq 0$ \\ 
        $\xi\neq(1-c^2)/6$
        \vspace{0.2cm} }\\
        \hline
        6 & \makecell{
        $\dfrac{-4\xi \left(1+c^2\right)}{4 \xi -1}$,\;\; 
        $-2$,\;\; 
        $\dfrac{ 3-4\xi ( 4+c^2 ) }{4 \xi -1}$,\;\; $\dfrac{3 c^2-2c^2 \xi ( 8+c^2 )-96 \xi ^2+34 \xi -3}{(4 \xi -1)B}$ \vspace{0.2cm}}
        & \makecell{stable, \vspace{0.1cm}\\ saddle \\ \\ } 
        & \makecell{ if $\xi<0$ or $1/4<\xi,\;\xi\neq 0$, \vspace{0.1cm}\\ if $0<\xi<1/4,\;\xi\neq 0$, \vspace{0.1cm}\\  }
        \\
        \hline
        7 & \makecell{ 
                \\
        $12 \xi -2 \sqrt{6\xi B},\;\;
        \left[ c^2 (3-12 \xi )+(6 \xi -1) \left(2\sqrt{6\xi B}-12 \xi +3\right)\right]/B,$\;\;
    \vspace{0.1cm}\\ 
        $\left[ 2 c^2 \left(\sqrt{6\xi B}-12 \xi +2\right)+2 (6 \xi -1) \left(2\sqrt{6\xi B}-12 \xi +2\right)\right]/B$,
    \vspace{0.1cm}\\ 
        $\left[ 2c^2 \left( \sqrt{6\xi  B}-12 \xi +3\right)+2(6 \xi -1) \left( 2\sqrt{6\xi B}-12 \xi +3\right)\right]/B$
    \vspace{0.2cm}
        } & 
        \makecell{
        \\saddle, 
        \vspace{0.2cm}\\ 
            unstable\\ spiral, 
        \vspace{0.2cm}\\ 
            unstable \\\\} & 
        \makecell{
            \vspace{0.1cm}\\ 
            if $\xi< 0,\;\xi\neq 0$, 
            \vspace{0.2cm}
            \\ 
            if $0<\xi<(-17+12c^2$\\$-\sqrt{1+24c^2})/[ 48(c^2-2)],\;\xi\neq 0  $,
            \vspace{0.2cm}
            \\
            if $(-17+12c^2$\\$-\sqrt{1+24c^2})/[ 48(c^2-2)< \xi ],\;\xi\neq 0 $
            \vspace{0.1cm}\\
         } 
         \\
        \hline
        8 & \makecell{\\ 
        $12 \xi +2 \sqrt{6\xi B},\;\;
        \left[ c^2 (3-12 \xi )+(6 \xi -1)\left(-2\sqrt{6\xi B}-12 \xi +3\right)\right]/B,$
        \vspace{0.1cm}\\ 
        $\left[ 2 c^2 \left(-\sqrt{6\xi B}-12 \xi +2\right)+2 (6 \xi -1) \left(-2\sqrt{6\xi B}-12 \xi +2\right)\right]/B$, 
        \vspace{0.1cm}\\ 
        $\left[ 2c^2 \left(-\sqrt{6\xi  B}-12 \xi +3\right)+2(6 \xi -1) \left(-2\sqrt{6\xi B}-12 \xi +3\right)\right]/B$
        \vspace{0.2cm}
        } & \makecell{\\
        unstable, 
        \vspace{0.2cm}\\ saddle \\
        spiral, 
        \vspace{0.2cm}\\ saddle \\ \\} & 
        \makecell{\vspace{0.1cm}\\ 
            if $\xi< 0,\;\xi\neq 0$,
       \vspace{0.2cm}\\ 
            if $0<\xi<(-17+12c^2$\\$-\sqrt{1+24c^2})/[ 48(c^2-2)],\;\xi\neq 0  $,
        \vspace{0.2cm}\\ 
       if $(-17+12c^2$\\$+\sqrt{1+24c^2})/[ 48(c^2-2)]<\xi,\;\xi\neq 0 $
        \vspace{0.2cm}\\} \\
        \hline
        9 & \makecell{\\
        $-3$,\;\; $-2$,\;\; 
        $\Big[3+3 \left(c^2-3\right) \xi-\sqrt{9 \left(c^2+5\right)^2 \xi ^2-6 \left(5 c^2+17\right) \xi +9} \Big] / ( 6\xi-2 )$,
        \vspace{0.1cm}\\ 
        $\Big[ 3+3 \left(c^2-3\right) \xi+\sqrt{9 \left(c^2+5\right)^2 \xi ^2-6 \left(5 c^2+17\right) \xi +9}\Big] / ( 6\xi-2 )$
        \vspace{0.2cm}
        } &\makecell{saddle, \\ stable} & \makecell { if $ \xi < 0 $ and $ 1/3<\xi ,\;\xi\neq 0 $, 
        \vspace{0.1cm}\\
        if $ 0<\xi<1/3,\;\xi\neq 0 $
        } \\
    \hline
    \bottomrule
    \end{tabular}
    \caption{Eigenvalues and their stabilities of all fixed points for $n=2$, $B=-1+c^2+6\xi$, $\Delta=c^4 [8 \xi  (194 \xi -23)+1]+32 c^2 \xi  (6 \xi -1) (116 \xi -13)+256 \xi  (9 \xi -1) (1-6 \xi )^2$, and $\Theta=3\xi [c^4 \xi  (48 \xi ^2+104 \xi -21)+6 c^2 (128 \xi ^3+40 \xi ^2-28 \xi +3)+2 (6 \xi -1) (3-16 \xi )^2]$
    }
    \label{table_eigenvalue}
\end{table}


\begin{table}
    \centering
    \renewcommand{\arraystretch}{2.5}
    \begin{tabular}{|c|c|}
    \hline
        Fixed points & equation of state parameter, $w_\text{eff}$  \\
    \hline\hline
        1 & \text{Indeterminate}  
        \\
    \hline
        2 & $-\dfrac{1}{3}$ 
        \\
    \hline
        3 & $-\dfrac{1}{3}$ 
        \\
    \hline
        4 & $-1$ 
        \\
    \hline
        5

        & $\dfrac{4\xi}{-3+12\xi}$ 
        \\
    \hline
        6 & $-1$ 
        \\
    \hline
        7 & $\dfrac{c^2(3-24\xi+2\sqrt{6\xi B})+(-1+6\xi)(3-24\xi+4\sqrt{6\xi B})}{3B}$ 
        \\
    \hline
        8 & $\dfrac{c^2(3-24\xi-2\sqrt{6\xi B})+(-1+6\xi)(3-24\xi-4\sqrt{6\xi B})}{3 B}$
        \\
    \hline
        9 & $-1$
        \\
    \hline
    \bottomrule
    \end{tabular}
    \caption{Effective equation of state of all fixed points for $n=2$ and $B=-1+c^2+6\xi$. This is found from the equation \eqref{eff_EOS} with the values of dynamical variables at the fixed points.}
    \label{tab:FP_eos}
\end{table}
\subsection{Fixed points and stability}

\subsubsection{Fixed point 1}
Fixed point 1 corresponds to a completely spatial curvature-dominated point, $\Omega_{k c}=1$. The effective equation of state parameter at this point is indeterminate due to the zero denominator in equation \eqref{eff_EOS}. The eigenvalues of the Jacobian matrix $\mathcal{M}$ (equation \eqref{Jacobian_matrix}) are given by, 
    \be
    \lambda_1=0,\quad \lambda_2=2,\quad \lambda_3=-1,\quad \lambda_4=-2.
    \ee
Since the eigenvalues include both positive and negative values, the stability of this point forms a saddle line for all $\xi$ and $0\leq c<1$, where the parameter $A_c\in(-\infty,\infty)$. However, due to the indeterminate value of $w_{\rm eff}$, this fixed point is considered non-physical.   

\subsubsection{Fixed point 2}
Fixed point 2 corresponds to a spatial curvature-dominated state, where the effective equation of state parameter is $w_\text{eff}=-{1}/{3}$. The eigenvalues of the Jacobian matrix are given by, 
    \begin{eqnarray}
    \nonumber
    &&\lambda_1 = 1,\quad \lambda_2 = 2,\\
    \nonumber
    &&\lambda_3 = \frac{-\left[4\xi \left(c^2-4\right) + c^2+96 \xi ^2\right]-\sqrt{ c^4 [8 \xi  (194 \xi -23)+1]+32 c^2 \xi  (6 \xi -1) (116 \xi -13)+256 \xi  (9 \xi -1) (1-6 \xi )^2}}{32 \xi (-1+c^2+6\xi)},\\
    &&\lambda_4 = \frac{-\left[4\xi \left(c^2-4\right) +c^2+96 \xi ^2\right]+\sqrt{c^4 \left[8 \xi  (194 \xi -23)+1\right]+32 c^2 \xi  (6 \xi -1) (116 \xi -13)+256 \xi  (9 \xi -1) (1-6 \xi )^2}}{32 \xi (-1+c^2+6\xi)}
    \end{eqnarray}
For $\xi < 0$ and $\xi>(4-3c^2)/24$, eigenvalues are mixed with positive and negative values hence it is a saddle point. The point is
unstable for $0 < \xi<(4-3c^2)/24$ which gives all positive eigenvalues. 
Singularity is at $\xi =0$ and $\xi = (1-c^2)/6$. The requirement for positivity of holographic density parameter is $\xi > 0$ however this makes $x_c < 0$ which is not physical. On the other hand, having  $\xi < 0$ makes $x_c > 0$, but this makes $\Omega_{\Lambda c} < 0$ which is also not physical. Hence this fixed point exists but it is non-physical for all range of $\xi$.

\subsubsection{Fixed point 3}

Fixed point 3 corresponds to a spatial curvature-dominated epoch, $\Omega_{kc}=1$, when   $\xi=(1-c^2)/6$. The equation of state parameter for this point is $w_\text{eff}=-1/3$ for all $\xi$.  The eigenvalues for this point are given by 
    \be
    \lambda_1=2,\quad \lambda_2=\frac{4(-1+c^2+6\xi)}{c^2},\quad \lambda_3=\frac{3c^2+24\xi-4}{c^2},\quad \lambda_4=\dfrac{6c^4\xi+4c^2(6\xi-1)+4(1-6\xi)^2}{3c^4\xi},
    \ee 
where $c \neq 0$ and $\xi \neq 0$ that is $0 < c < 1$. 
According to the eigenvalues, for $\xi<(8-c^4-4c^2+c^3\sqrt{8+c^2})/48$, the fixed point is a saddle point.   For $\xi\geq(8-c^4-4c^2+c^3\sqrt{8+c^2})/48$, the fixed point is unstable point. 


\subsubsection{Fixed point 4}
Fixed point 4 corresponds to $w_\text{eff}=-1$. The eigenvalues are given by, 
    \begin{eqnarray}
    \nonumber
    &&\lambda_1=-2,\quad \lambda_2=3,\\ 
    \nonumber
    &&\lambda_3 = \frac{3 c^2  (4 \xi -1)-\sqrt{3\xi \left[ c^4 \xi  \left(48 \xi ^2+104 \xi -21\right)+6 c^2 \left(128 \xi ^3+40 \xi ^2-28 \xi +3\right)+2 (6 \xi -1) (3-16 \xi )^2\right]}}{4( -1+c^2+6\xi )} 
    \\
    &&\lambda_4 = \frac{3 c^2  (4 \xi -1) + \sqrt{3\xi \left[c^4 \xi  \left(48 \xi ^2+104 \xi -21\right)+6 c^2 \left(128 \xi ^3+40 \xi ^2-28 \xi +3\right)+2 (6 \xi -1) (3-16 \xi )^2\right]}}{4( -1+c^2+6\xi )}.
    \end{eqnarray} 
Since there is one positive eigenvalue, this fixed point is saddle for $0\leq c<1$. Requirement of $0\leq\Omega_{\m c}\leq1$ imposes a condition, $
    {3}/({20+4c^2}) \leq \xi \leq {3}/({16+4c^2})\,$ which implies positive value of $\xi$ in the range. For it to be physically valid, the condition $x_c > 0$ and $y_c > 0$ is required and these lead to $\xi < 0$.
    The conditions are in conflict with each other, hence this point is non-physical.  

\subsubsection{Fixed point 5}
\begin{figure}
    \centering
    \includegraphics[width=6cm]{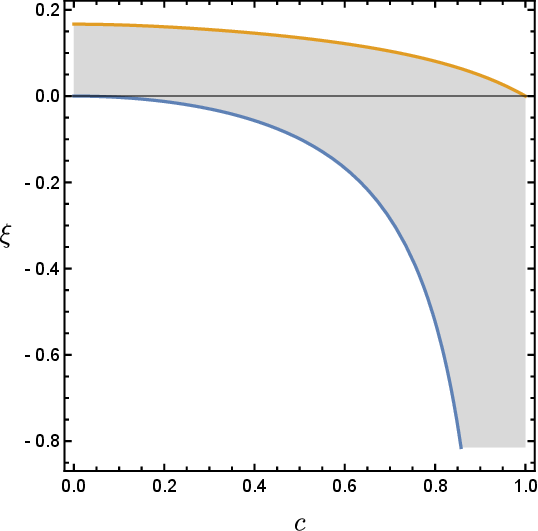}
    \caption{Parametric plot between $c$ and $\xi$ for the fixed point 5 shows the allowed region of $\xi$. In addition, the condition $x_c \geq 0$ requires $\xi < 0$ corresponding to the region above the blue curve but below $\xi = 0$.}
    \label{fp5paraplot}
\end{figure}

 The eigenvalues of this fixed point depend on $\xi$ only when $\xi \neq 0$. These are given by 
     \be
    \lambda_1 = \frac{4\xi}{1-4\xi},\quad \lambda_2=\frac{1-8\xi}{1-4\xi},\quad \lambda_3=\frac{3-16\xi}{1-4\xi},\quad \lambda_4=\frac{3c^2(1-4\xi)^2-96\xi^2+34\xi-3}{ 2( 4\xi-1 )( -1+c^2+6\xi )},
    \ee 
where $\xi\neq 1/4$ and $\xi\neq(1-c^2)/6$. Since $x_c$ must be positive and this corresponds to $\xi < 0$, the eigenvalues are either positive or negative. Therefore this fixed point is a saddle point.   This fixed point corresponds to \be w_{\rm eff} = \dfrac{4\xi}{-3+12\xi} \ee which is independent of $c$. Acceleration condition $w_{\rm eff} < -1/3$ permits a range of the NMC coupling, $1/8 < \xi < 1/4$.   The dust-like equation of state, $w_{\rm eff} = 0$, requires $\xi \rightarrow 0$.  The equation of state, $w_{\rm eff} = -1$, requires $\xi = 3/16$.  
Condition of $x_c < 0$ requires $\xi \leq 0$, hence this fixed point does not result in acceleration nor dust-like expansion and nor (effectively) de-Sitter expansion.
As seen in TABLE \ref{table_fixedpoint}, the condition $0\leq\Omega_{{\rm m}0}\leq1$, implies 
    \be
        \frac{17-12c^2+\sqrt{1+24c^2}}{96-48c^2}\:  \leq \: \xi \: \leq \: \frac{5-12c^2+\sqrt{25+24c^2}}{48-48c^2}\,,
        \label{fp5a}
    \ee
and 
    \be
    \frac{-5+12c^2+\sqrt{25+24c^2}}{-48+48c^2}\: \leq \: \xi \: \leq \: \frac{-17+12c^2+\sqrt{1+24c^2}}{-96+48c^2}\,.
    \label{fp5b}
    \ee  
    Both conditions depend on $c$. The condition $x_c > 0$ requires that $\xi < 0$, allowing only the condition \eqref{fp5b}. Parametric plot of the condition \eqref{fp5b} is shown in Fig. \ref{fp5paraplot} where the allowed region for $\xi$ to be valid is the shaded area above the blue curve and under the line $\xi = 0$. 
         As in TABLE \ref{table_fixedpoint},  for dust matter to be completely dominating, 
\be
    \Omega_{{\rm m}c}\,=\, -c^2+\dfrac{3-34\xi+96\xi^2}{3(1-4\xi)^2} = 1\,,
\ee
it follows that as $\xi \rightarrow 0$ with $\Omega_{{\rm m}c} \rightarrow  1$ directly gives $c \rightarrow 0$. Canonical scalar field in GR limit is approached when  $\xi \rightarrow 0$ and $c \rightarrow  0$. 
This fixed point does not completely recover the Canonical scalar GR limit as $\xi$ must be negative. Approaching the limit, $\xi\rightarrow 0^{-}$ and $c  \rightarrow 0^{-}$ results in $\Omega_{{\rm m}c} \rightarrow  1^{-}$. This point approaches 
matter-domination limit as $\xi \rightarrow 0$ and it is a saddle point transition.

\subsubsection{Fixed point 6}
Eigenvalues of the Jacobian matrix at this point are given by 
\be
    \lambda_1 = \frac{-4\xi( 1+c^2 )}{ 4\xi-1 },\quad 
    \lambda_2 = -2,\quad 
    \lambda_3 = \dfrac{ 3-4\xi ( 4+c^2 ) }{4 \xi -1},\quad 
    \lambda_4 = \dfrac{3 c^2-2c^2 \xi ( 8+c^2 )-96 \xi ^2+34 \xi -3}{(4 \xi -1)( -1+c^2+6\xi )},
    \ee 
where $\xi \neq 1/4$ and $\xi\neq(1-c^2)/6$.  The stability therefore depends only on $\xi$ and this can be considered in three cases. First, for $0<\xi<1/4$, the fixed point is saddle. Second, for $\xi< 0$ or $1/4< \xi$, the fixed point is stable. Third, for $\xi=0$, the system becomes indeterminate due to zero denominators in equations \eqref{x,s,A_constraint}, \eqref{Friedmann_constraint_2}, 
\eqref{eq: autonomous system}. To study this model in the canonical scalar holographic limit ($\xi = 0$), one must redefine the dynamical variables such that it is a completely new autonomous system and there is no indeterminate term. Here, this case is considered as approaching limit only.   
Generally, the kinetic term must not be negative and the condition $x_c \geq 0$ must hold. As a result, this requires $\xi < 0$ for validity.
    This fixed point corresponds to $w_\text{eff}=-1$ for all $\xi$ and $0\leq c<1$. 
NMC scalar field and the holographic vacuum energy contribute to the dark energy with $\Omega_{\text{DE} c}=\Omega_{\phi c}+\Omega_{\Lambda c}=x_c+y_c+s_c+\Omega_{\Lambda c}=1$, implying dark energy domination. 
Considering the canonical scalar limit of the GR theory, $\xi \rightarrow 0, (\xi \neq 0) $ and $c \rightarrow 0$, the potential term is dominant at late time, i.e. $x_c \rightarrow 0, y_{c} \rightarrow 1$ (the potential acts solely as dark energy). For $\xi \rightarrow 0$ and $c \neq 0$, we have $x_c \rightarrow 0, y_{c} \rightarrow 1 - c^2$ and $\Omega_{\Lambda c} = c^2$. For $\xi \neq 0$ and $c = 0$, the results agree with the results previously reported by Sami {\it et al.} \cite{Sami:2012uh} for the non-holographic NMC theory.

\subsubsection{Fixed point 7}
Eigenvalues of this fixed point are given by 
    \begin{eqnarray}
        \no&&\lambda_1 = 12 \xi -2 \sqrt{6\xi (-1+c^2+6\xi)},
        \\
        \no&&\lambda_2 = \frac{ c^2 (3-12 \xi )+(6 \xi -1) \left[ 2\sqrt{6\xi (-1+c^2+6\xi)}-12 \xi +3\right] }{ -1+c^2+6\xi },   
        \\
        \no&&\lambda_3 = \frac{ 2 c^2 \left[ \sqrt{6\xi (-1+c^2+6\xi)}-12 \xi +2\right] + 2 (6 \xi -1) \left[ 2\sqrt{6\xi (-1+c^2+6\xi)}-12 \xi +2\right] }{ -1+c^2+6\xi },
        \\
        &&\lambda_4 = \frac{ 2c^2 \left[ \sqrt{6\xi (-1+c^2+6\xi)}-12 \xi +3 \right]+2(6 \xi -1) \left[ 2\sqrt{6\xi (-1+c^2+6\xi)}-12 \xi +3\right] }{ -1+c^2+6\xi }\,.
    \end{eqnarray}
To avoid divergence in the eigenvalues, the coupling parameter must satisfy $\xi\neq{(1-c^2)}/{6}$.   
As in TABLE \ref{tab:FP_eos}, the equation of state for this point is,
\be
w_{\rm eff} = \dfrac{c^2\left[ 3-24\xi+2\sqrt{6\xi(-1+c^2+6\xi)} \right]+(-1+6\xi)\left[ 3-24\xi+4\sqrt{6\xi(-1+c^2+6\xi)}\right] }{3(-1+c^2+6\xi)}\,.
\ee
For $\xi< 0$, and $0\leq c<1$, the fixed point exhibits saddle point stability. For 
$\xi \rightarrow 0^-$ and $0\leq c<1$, the fixed point approached canonical scalar holographic case (without the NMC effect). This gives $w_{\rm eff} \rightarrow 1^-$ which effectively is the stiff fluid.   
For $\xi>0$, the stability depends on the range of $\xi$ as follows: if $0<\xi<(-17+12c^2-\sqrt{1+24c^2})/[48(c^2-2)]$, the eigenvalues are unstable spiral; if $\xi>(-17+12c^2-\sqrt{1+24c^2})/[48(c^2-2)]$, the point is unstable point. 
This is however, condition of the kinetic term, $x_c \geq 0$ with a condition $0 \leq c < 1$, implies $\xi < 0$ (see TABLE \ref{table_fixedpoint}) so that the point can be physical.

In the canonical scalar GR limit, $\xi\rightarrow 0^{-}$ and $c\rightarrow 0^{+}$, this point is a purely kinetic-dominated fixed point with $x_c=1$ and \be
\lim_{c\rightarrow 0^{+}, \xi\rightarrow 0^{-}} \: w_{\rm eff} = 1\,,
\ee
which corresponds to a stiff fluid-dominated epoch.

\subsubsection{Fixed point 8}

Eigenvalues of this fixed point are given by 
    \begin{eqnarray}
        \no&&\lambda_1 = 12 \xi + 2 \sqrt{6\xi (-1+c^2+6\xi)},
        \\
        \no&&\lambda_2 = \frac{ c^2 (3-12 \xi )+(6 \xi -1) \left[ -2\sqrt{6\xi (-1+c^2+6\xi)} - 12 \xi +3\right]}{ -1+c^2+6\xi },   
        \\
        \no&&\lambda_3 = \frac{ 2 c^2 \left[ -\sqrt{6\xi (-1+c^2+6\xi)}-12 \xi +2\right] + 2 (6 \xi -1) \left[ -2\sqrt{6\xi (-1+c^2+6\xi)}-12 \xi +2\right]}{ -1+c^2+6\xi },
        \\
        &&\lambda_4 = \frac{ 2c^2 \left[ -\sqrt{6\xi (-1+c^2+6\xi)}-12 \xi +3 \right]+2(6 \xi -1) \left[ -2\sqrt{6\xi (-1+c^2+6\xi)}-12 \xi +3\right]}{ -1+c^2+6\xi }\,.
    \end{eqnarray}
and the equation of state is
\be
w_{\rm eff} = \dfrac{c^2\l(3-24\xi-2\sqrt{6\xi(-1+c^2+6\xi)}\r)+(-1+6\xi)\l(3-24\xi-4\sqrt{6\xi(-1+c^2+6\xi)}\r)}{3(-1+c^2+6\xi)}\,.
\ee
Under the limit $\xi \rightarrow 0^-$ and $0 \leq c < 1$, this fixed point is similar to  the fixed point 7, where there are only kinetic term and constant vacuum energy in the universe. This results in stiff-fluid equation of state parameter $w_\text{eff}=1$.
Stiff fluid can also appear in the canonical scalar GR limit, $\xi\rightarrow 0^{-}$ and $c\rightarrow 0^{+}$, as $
\lim_{c\rightarrow 0^{+}, \xi\rightarrow 0^{-}} \: w_{\rm eff} = 1\,.$
The stability can be analyzed in three cases as follows: 
For $\xi<0$, the fixed point is a unstable point for $0\leq c<1$. For $\xi>0$, stability depends on the range of $\xi$: if $0<\xi<(-17+12c^2-\sqrt{1+24c^2})/[48(c^2-2)]$, the point is saddle spiral; if $\xi>(-17+12c^2-\sqrt{1+24c^2})/[48(c^2-2)]$, the stability is a saddle point.
The condition $x_c \geq 0$ with $0 \leq c < 1$ forces that $\xi < 0$ for it to be physically valid.

\subsubsection{Fixed point 9}
Fixed point 9 corresponds to 
constant vacuum energy and potential domination with $w_\text{eff}=-1$. The eigenvalues for this fixed point are
    \begin{eqnarray}
    \no&&\lambda_1 = -3,\quad \lambda_2=-2,
    \\ 
    \no&&\lambda_3 = \frac{ 3+3 \left(c^2-3\right) \xi-\sqrt{9 \left(c^2+5\right)^2 \xi ^2-6 \left(5 c^2+17\right) \xi +9} }{ 6\xi-2 },
    \\
    \no&&\lambda_4 =  \frac{ 3+3 \left(c^2-3\right) \xi+\sqrt{9 \left(c^2+5\right)^2 \xi ^2-6 \left(5 c^2+17\right) \xi +9} }{  6\xi-2 }
    \end{eqnarray}
    For $\xi < 0$ or $1/3 < \xi $, the point is saddle and for $0 < \xi < 1/3$ the point is stable. The NMC coupling must not be $0$ nor $1/3$, ($\xi \neq 0, 1/3$). This point is potential and vacuum energy dominated as seen in TABLE \ref{table_fixedpoint}. It corresponds to $w_{\rm eff} = -1$.


 \begin{figure}
    \centering
    \includegraphics[width=0.50\linewidth]{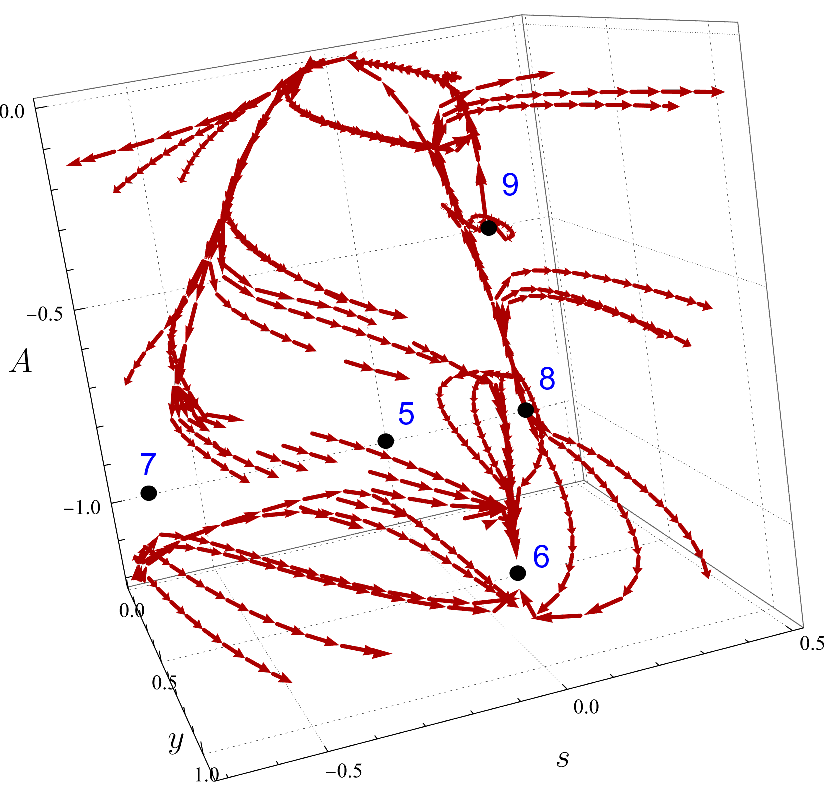}
    \\
    \includegraphics[width=0.45\linewidth]{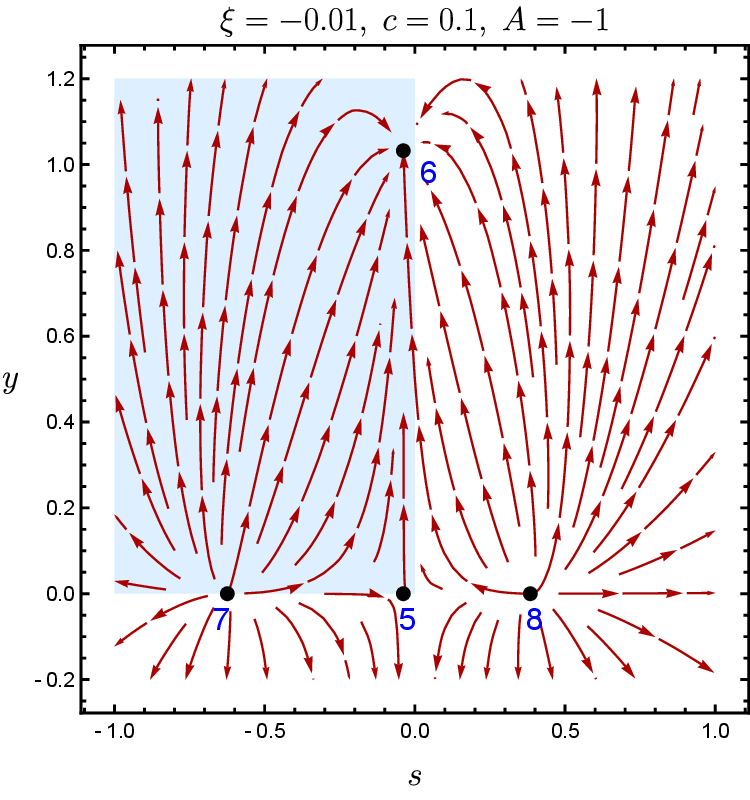}\hspace{0.5cm}
    \includegraphics[width=0.45\linewidth]{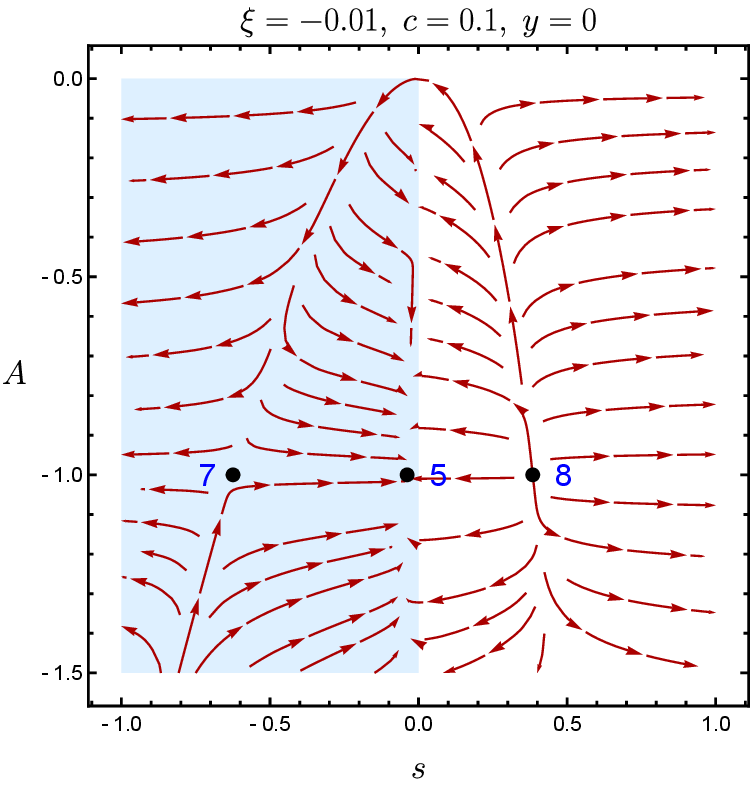}
    \caption{Phase portrait illustrates the evolution of a flat universe $(\Omega_k=0)$ with fixed parameters $c=0.1$ and $\xi=-0.01$. The upper figure shows a three-dimensional phase portrait of variables: $y,s$ and $A$. The bottom-left panel presents the phase plot on the $ys-$plane at $A=-1$, while the bottom-right panel presents the phase plot on $As-$plane at $y=0$. Shading areas are allowed valid regions of the model.}
    \label{fig:phase_plot}
\end{figure}

\section{Phase portrait} \label{ppt}

Observational data of the current curvature density parameter from DESI + CMB + Union3 \cite{DESI2024} indicates a small value of about $\Omega_{k0} = -0.0004 \pm 0.0019$. Hence the present universe is very close to flatness, and we assume flat space here. According to TABLE \ref{table_eigenvalue}, dynamical system analysis identifies three fixed points (points 1, 2 and 3) corresponding to non-flat cases and six fixed points (points 4, 5, 6, 7, 8 and 9) corresponding to flat cases. Here we discuss the flat cases which are of points 4 to 9.

One can think of a universe, omitting a radiation-dominated epoch, that begins with an almost stiff fluid-domination corresponding to the fixed point 7 or 8\footnote{For the point 7 and 8, according to TABLE \eqref{tab:FP_eos}, as $\xi \rightarrow 0^{-}$, the effective equation of state approaches stiff-fluid condition, $w_{\rm eff} \rightarrow 1^-$.}. 
The physical condition $x_c \geq 0$ requires that $\xi<0$ and it is required that $\xi \neq 0$ for the theory to be valid. In addition, condition $\xi \rightarrow 0^-$ is required for, first, to satisfy the dust domination condition approaching the saddle fixed point 5, and second, to achieve the kinetic dominated (stiff fluid) condition of the fixed points 7 and 8. We can conclude that this model requires $\xi \rightarrow 0^-$ to be physically valid.   With these conditions, the dynamical variable, 
$s={16\pi G_\text{eff} \xi \phi \dot{\phi} }/{H}$, 
must be negative for an expanding universe ($H > 0$). 
According to TABLE \ref{table_eigenvalue}, for $\xi\rightarrow 0^-$, the fixed point 7 is a saddle, while the fixed point 8 is an unstable point. For $\xi\rightarrow 0^-$, the fixed point 8 locates in $s > 0$ region which does not match the expanding universe, since it needs $H < 0$ (considering positive field value).  In addition, for the condition $\xi \rightarrow 0^-$, the variables 
 $y = 8\pi G_\text{eff} V_0 \phi^2/{3H^2} $ 
 and 
$A=8\pi G_\text{eff}\xi \phi^2$ need to take condition $0 < y$ and $A < 0$. 
Our interest is therefore to choose the fixed point 7 which locates in the allowed region ($s<0, A<0$ and $y>0$) as the beginning of the cosmic evolution.
This represents a universe beginning with scalar kinetic term and holographic vacuum energy domination (almost stiff-fluid equation of state: the saddle point 7) transiting into an almost-dust dominate universe (the saddle point 5)\footnote{For the fixed point 5, according to TABLE \eqref{tab:FP_eos}, there is no completely dust-dominated epoch ($w_\text{eff}
=0$) however as $\xi \rightarrow 0^{-}$, this approach dust-dominated condition.}. After leaving the saddle point 5, the universe evolves to the stable point 6 which is a dark energy dominated universe at late time.    
The path of evolution are concluded as $7 \rightarrow 5 \rightarrow 6$. FIG \ref{fig:phase_plot} shows three-dimensional phase portrait of a space $(s, y, A)$.        Bottom panels of the figure are two plane slices. The left one is $ys-$plane 
with $\xi=-0.01$, $c=0.1$ and $A=-1$ and the right one is  $As-$plane 
with $\xi=-0.01$, $c=0.1$ and $y=0$. Fixed points are labeled in the figures. Regions in shading are the allowed regions. In these figures, trajectories from $7 \rightarrow 5$ can be seen and at last the phase portraits evolve to the stable fixed point 6. This approximately corresponds to $ y_c \rightarrow 1.02,\;\;s_c \rightarrow -0.038$. The fixed point 6 represents a stable cosmological constant-dominated epoch $w_\text{eff}=-1$, which is consistent with current observations.  Although fixed points 4 and 9 also correspond to dark energy domination, they are excluded from our analysis as they are not stable under the condition $\xi<0$.


 Here the dark energy is effective contribution of both scalar field and holographic vacuum energy. 
 From the Friedmann constraint, \eqref{FM_constraint}, the stable fixed point 6 satisfies $
 1 = x_c + y_c + s_c +\Omega_{{\rm m}c}+\Omega_{kc}+\Omega_{\Lambda c},
 $
 with $\Omega_{{\rm DE}c} = x_c + y_c + s_c + \Omega_{\Lambda c} =        \Omega_{\phi c}+\Omega_{\Lambda c} = 1$.

\begin{figure}
    \centering
    \includegraphics[width=0.45\linewidth]{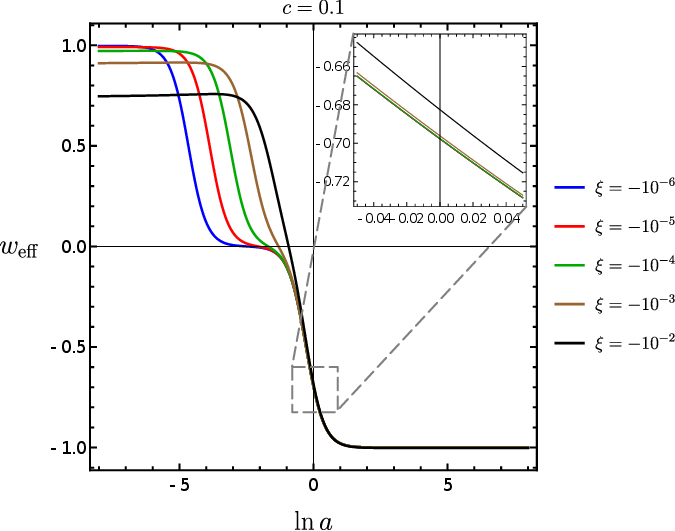}
    \includegraphics[width=0.43\linewidth]{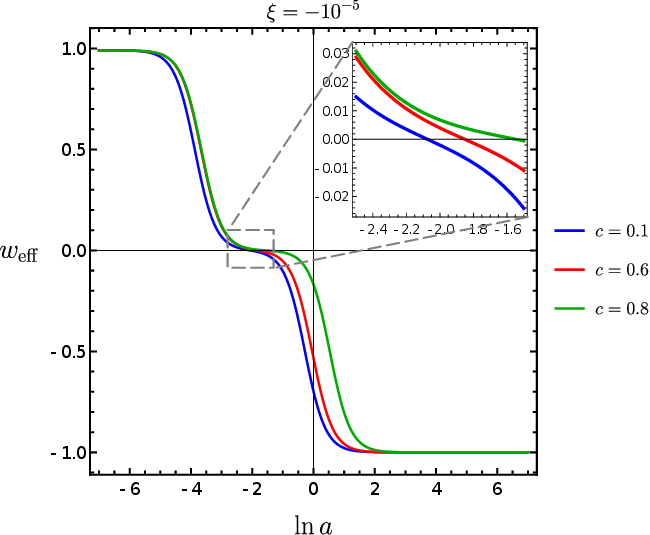}
    
    \caption{The figure shows evolution of the effective equation of state parameter $w_\text{eff}$ by $\ln a$. In this plot, consistent initial conditions are  $s_0=-10^{-7},\;\;A_0=-0.7,\;\;\Omega_{k0}=-0.0004$,  $\Omega_{\text{m}0}=0.3233$ and  $y_0=1-s_0-\Omega_{\text{m}0}-s_0^2/(24 A_0 \xi)-c^2(1-\Omega_{k0})-\Omega_{k0}$. Larger magnitude of negative NMC coupling results in less and less effect of dust matter domination. Holographic effect $c$ increases the value of EoS.}
    \label{fig:weff_vary_c}
\end{figure}

\begin{figure}
    \centering
    \includegraphics[width=0.30\linewidth]{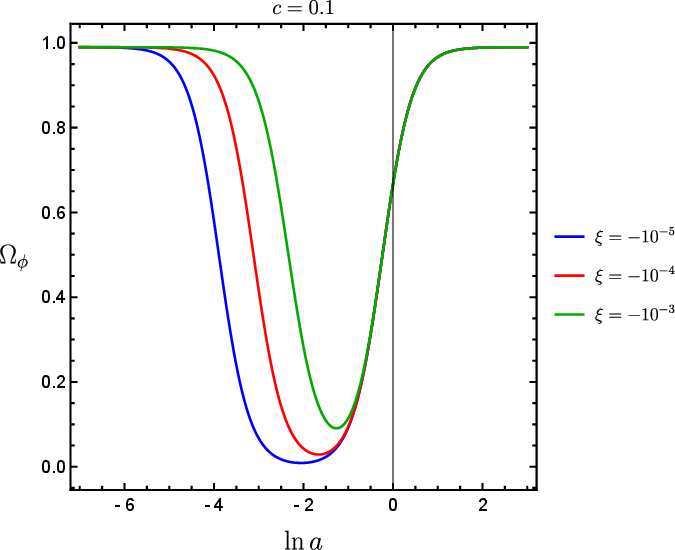} 
    \includegraphics[width=0.33\linewidth]{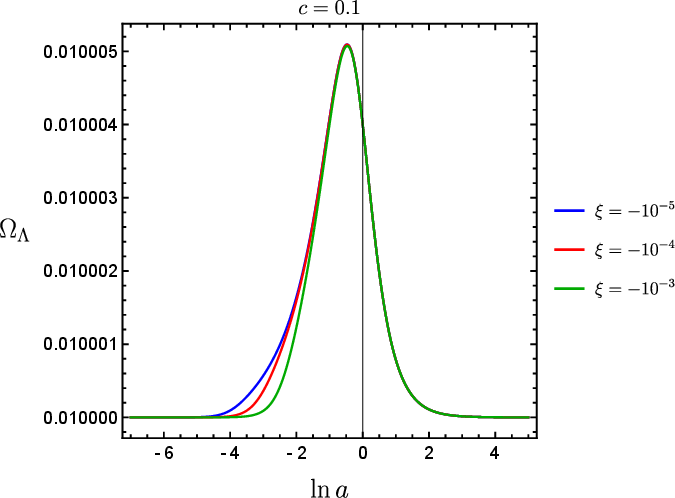} 
    \includegraphics[width=0.32\linewidth]{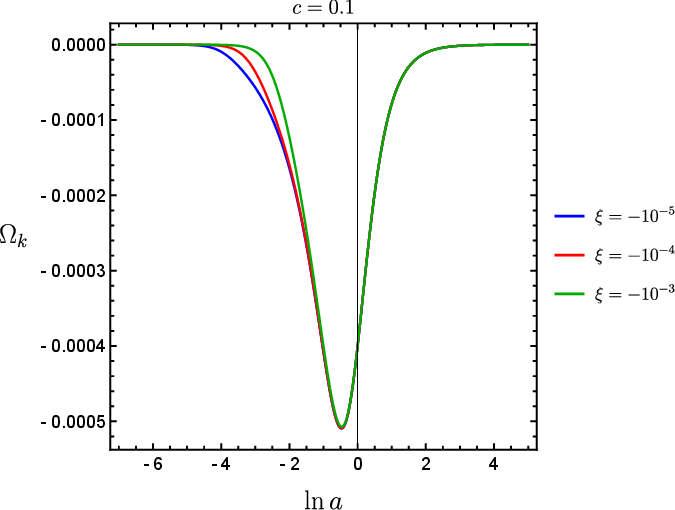} 
    \includegraphics[width=0.30\linewidth]{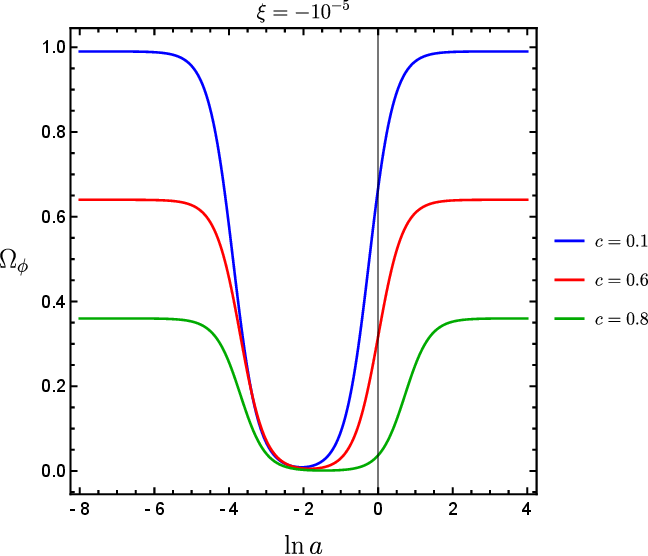} 
    \includegraphics[width=0.30\linewidth]{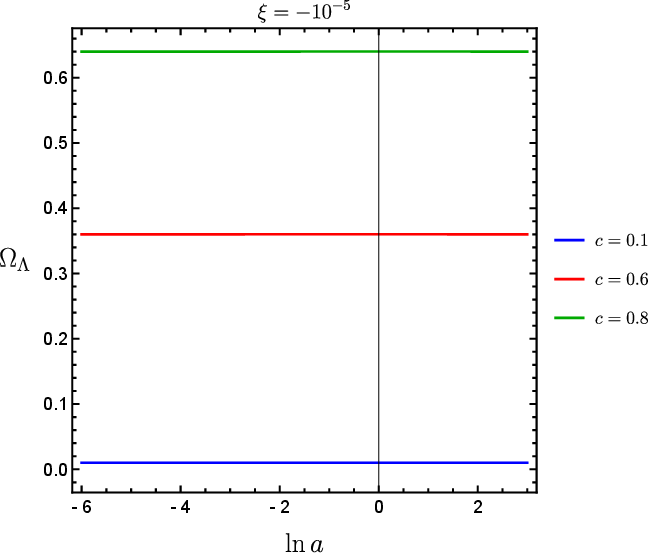} 
    \includegraphics[width=0.32\linewidth]{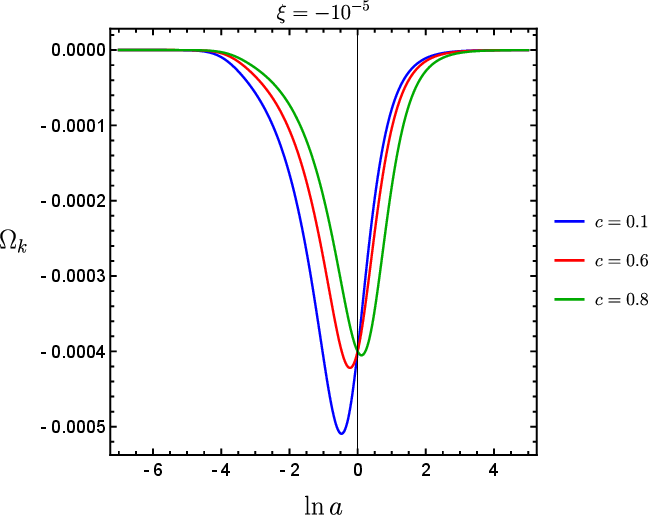}
    \caption{The figure shows evolution of the NMC scalar field density parameter $\Omega_\phi$, holographic vacuum energy density parameter $\Omega_\Lambda$ and spatial curvature density parameter $\Omega_k$ by $\ln a$. In this plot, consistent initial conditions are  $s_0=-10^{-7},\;\;A_0=-0.7,\;\;\Omega_{k0}=-0.0004$, $\Omega_{\text{m}0}=0.3233$ and $y_0=1-s_0-\Omega_{\text{m}0}-s_0^2/(24 A_0 \xi)-c^2(1-\Omega_{k0})-\Omega_{k0}$.}
    \label{fig:scalar_density_plot}
\end{figure}
\begin{figure}
    \centering
    \includegraphics[width=0.31\linewidth]{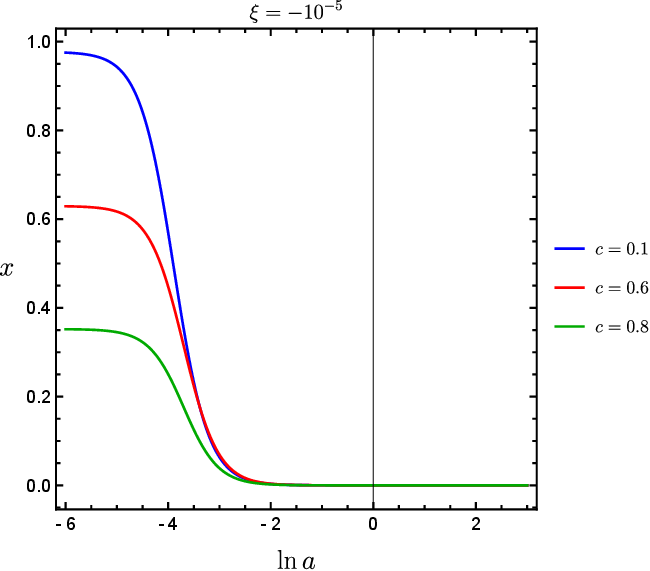} 
    \includegraphics[width=0.31\linewidth]{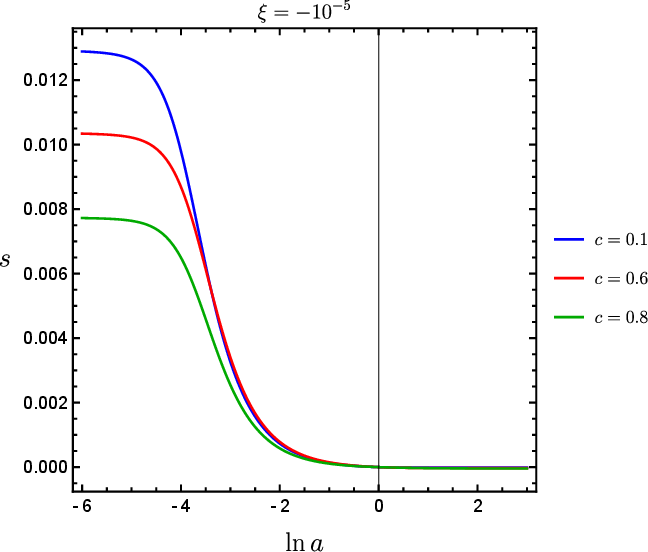} 
    \includegraphics[width=0.31\linewidth]{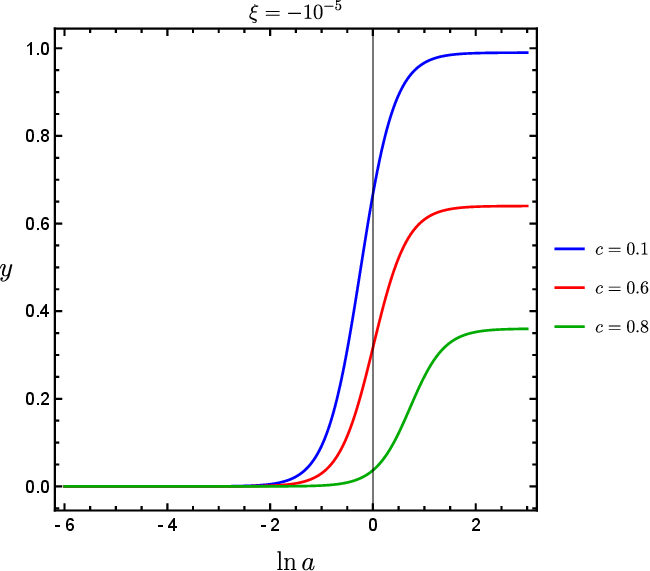}
    
    \caption{The figure shows evolution of variables $x,s,y$.
    In this plot, consistent initial conditions are  $s_0=-10^{-7},\;\;A_0=-0.7,\;\;\Omega_{k0}=-0.0004$,  $\Omega_{\text{m}0}=0.3233$ and $y_0=1-s_0-\Omega_{\text{m}0}-s_0^2/(24 A_0 \xi)-c^2(1-\Omega_{k0})-\Omega_{k0}$.}
    \label{fig:scalar_plot}
\end{figure}

%
%
\section{Numerical solutions} \label{num}

The autonomous system \eqref{eq: autonomous system} can be integrated numerically. The effective equation of state parameter, $w_\text{eff}(N)$, dust density parameter, $\Omega_\text{m}(N)$, dark energy density parameter, $\Omega_\text{DE}(N)$, and spatial curvature density parameter, $\Omega_k(\ln a)$, are plotted with $N=\ln a$. With quadratic potential $V(\phi)=V_0\phi^2$, FIG. \ref{fig:weff_vary_c} shows the effective equation of state parameter with varying holographic parameter $c=0.1,\,c=0.6,\,c=0.8$ (right panel) and coupling parameter $\xi=-10^{-6},\,\xi=-10^{-5},\,\xi=-10^{-4},\,\xi=-10^{-3},\,\xi=-10^{-2}$ (left panel). We choose consistent initial conditions as follows: $s_0=-10^{-7},\;A_0=-0.7$. The initial condition for parameter $y_0$ is given by Friedmann constraint equation \eqref{Friedmann_constraint_2} with observational value,  $\Omega_{\text{m}0}=0.3233$ and $\Omega_{k0}=-0.0004$ (DESI+CMB+Union3 ($w_0w_a$CDM+$\Omega_k$ model) \cite{DESI2024}). Hence $y_0=1-s_0-\Omega_{\text{m}0}-s_0^2/(24 A_0 \xi)-c^2(1-\Omega_{k0})-\Omega_{k0}$
where the values of $s_0$ and $A_0$ are chosen by hand.  According to TABLE \ref{tab:FP_eos}, no fixed points are completely dominated by dust or stiff fluids. For the cosmic evolution to approach dust or stiff fluid dominations, the coupling parameter must be small and negative ($\xi\rightarrow 0^-$), and the holographic parameter must also be small ($c\rightarrow 0$). FIG. \ref{fig:weff_vary_c} shows that, for any values of $\xi$ and $c$, the equation of state parameter $w_\text{eff}$ approaches $-1$ at late times, corresponding to a cosmological constant-like behavior. As the coupling parameter approaches a small negative value (e.g. $\xi=-10^{-6}$), the evolution of $w_{\rm eff}$ follows the phase portrait path $7\rightarrow 5 \rightarrow 6$. When the coupling parameter decreases further (i.e. its negative magnitude is larger, e.g. $\xi=-10^{-2}$), the dust-dominated EoS gradually disappears. This corresponds to increasing $w_\text{eff}$ values at the fixed points 5 and decreasing of $w_\text{eff}$ values at the fixed point 7 as seen in FIG. \ref{fig:weff_vary_c} 
 and TABLE \ref{tab:FP_eos}. In the right column of FIG. \ref{fig:weff_vary_c}, increasing of the parameter $c$ does not change the overall shape of the $w_{\rm eff}$ evolution. However, larger $c$ increases values of the equation of state parameter. For $\xi = -10^{-5}$  and   $c = 0.8$  the current universe does not undergo accelerated expansion, as $w_{\rm eff}>-1/3$. In conclusion, very small negative magnitude of $\xi$ and very small $0<c \ll 1$ are preferred in this model where late-time acceleration is driven by both scalar field and holographic vacuum energy.

FIG. \ref{fig:scalar_density_plot} illustrates the density parameters of the scalar field and holographic vacuum energy and the spatial curvature density parameter for 
the phase portrait path $7\rightarrow 5 \rightarrow 6$.
Top row of  FIG. \ref{fig:scalar_density_plot} shows evolution of these parameters with varying $\xi$ and bottom row presents evolution of these parameters with varying $c$.  As seen in FIG. \ref{fig:scalar_density_plot},  the coupling parameter $\xi$ does not significantly affect $\Omega_\Lambda$, whereas the parameter $c$ increases the value of $\Omega_\Lambda$, keeping it nearly constant. This occurs because $\Omega_k$ is very small, which, according to equation \eqref{holographic_constraint}, leads to a nearly constant $\Omega_\Lambda$. Evolution of $\Omega_k$ is in the right column of this figure.

Density parameter of the NMC scalar field is given by 
    $ \Omega_\phi=x+y+s, $
and its evolution is shown in the first column of FIG. \ref{fig:scalar_density_plot}. As the magnitude of negative coupling constant is larger,  the value of  $ \Omega_\phi$ is pushed up in the past.  However larger $c$ reduces the value of $ \Omega_\phi$.
Three terms $x, y, s$ that contribute to $ \Omega_\phi$ are plotted in  FIG. \ref{fig:scalar_plot}.  Kinetic term $x$, NMC term $s$ and potential term $y$
 are weaken for larger $c$. 
The variable $s$ has small effect during this period, as seen in the middle column of FIG. \ref{fig:scalar_plot}. In the right column of FIG. \ref{fig:scalar_plot}, the potential term $y$ becomes dominant at late times.

\section{Conclusions} \label{conclusion}

In this work, we consider 
non-minimally coupling theory (NMC) which is a subclass of the Horndeski theory. Previous study shows that the NMC theory with quadratic potential is viable for $\xi > -7.0 \times 10^{-3}$ 
\cite{Tsujikawa:2004my}. Ideas of Holographic Dark Energy (HDE) are also incorporated in our consideration. In HDE ideas, vacuum energy density is sum of zero-point one-loop correction of energy density and this is given by the CKN bound as  
$
\rho_{\Lambda} = {3c^2}/{8\pi G L^2} $ where $L$ is infrared cutoff scale and $0 \leq c<1$.
We consider FRW universe containing NMC scalar field, dust matter and holographic vacuum energy.  
In this model, effective gravitational constant is function of NMC scalar field, expressed as $G_\text{eff}(\phi)$ which is defined naturally at the action level. 
Therefore, with effective gravitational constant, the universe is filled with canonical scalar field, dust matter,  
and the holographic vacuum energy density. Therefore, in the definition of holographic vacuum density, we use effective gravitational constant, i.e.     
$
 \rho_{\Lambda} = {3c^{2}}/{8\pi G_{\text{eff}}L^{2}}\,.
 $
 The cutoff scale is chosen to be apparent horizon since it corresponds to a trapped null surface in similar sense to blackhole's interior surface. Therefore it is a natural choice for the holographic cutoff. We consider quadratic power-law scalar potential, $V(\phi) = V_{0}\phi^{2}$ here. Our dynamical system analysis of the four independent dimensionless parameters reveals nine fixed points as shown in TABLE \ref{table_fixedpoint}. In the definition of dynamical variables, there are directly NMC effects via $G_\text{eff}$ in all dynamical variables except in $\Omega_k$.  The holographic effect appears in $\Omega_\Lambda$ only. 
Both holographic vacuum energy and the scalar field are added up to dark energy density driving the acceleration. As in TABLE \ref{table_eigenvalue}, three non-physical fixed
points are the points 1, 2 and 4. The fixed point 3 corresponds to curvature term domination which can be either saddle or unstable. 
Fixed point 5 requires $\xi < 0$ and it is a saddle point. 
As $\xi \rightarrow 0$ and $c \rightarrow 0$, it approaches  
canonical scalar field theory in GR limit.  
Hence it is not a complete matter-domination but only a transient event in the evolution and this happens as $\xi \rightarrow 0^-$.  
Fixed point 6 corresponds to $w_\text{eff}=-1$ for all $\xi$ and $0\leq c<1$. The value range of  $\xi$ affects its stability directly. For $0<\xi<1/4$, the point is saddle.  
 For $\xi< 0$ or $1/4< \xi$, the point is stable. The case $\xi=0$ is not allowed since the system becomes indeterminate.  
We can not say that $\xi = 0$ is the canonical scalar holographic limit. This is because the dynamical variables must be re-defined from the beginning as such the system  becomes completely different autonomous system and it is not our interest here.
Generally, the condition $\xi < 0$ is favored
and in the non-holographic limit, we recover the NMC results reported previously by  Sami {\it et al.} \cite{Sami:2012uh}.
Fixed point 7 requires $\xi< 0$ and $0\leq c<1$ and it is a saddle point. In the limit 
$\xi \rightarrow 0^-$ and $0\leq c<1$, then $w_{\rm eff} \rightarrow 1^-$, i.e. the fixed point 7 corresponds to stiff-fluid domination.   
 The fixed point 8 also corresponds to stiff-fluid domination in the canonical scalar GR limit, $\xi\rightarrow 0^{-}$ and $c\rightarrow 0^{+}$.
Stability of the fixed point 8 comes in three cases.
For $\xi<0$, it is unstable otherwise they are either saddle spiral or saddle point.
Nevertheless, it is only physically valid when $\xi < 0$. 
Fixed point 9, depending on value of $c$, corresponds to constant vacuum energy and potential domination with $w_\text{eff}=-1$. 
    The point is saddle for $\xi < 0$ or $1/3 < \xi$ and is stable for $0 < \xi < 1/3$. Singularities are at $\xi =0, 1/3$.       

We consider only flat case here. In TABLE \ref{table_eigenvalue}, three fixed points (points 1, 2 and 3) are non-flat cases and six fixed points (points 4, 5, 6, 7, 8 and 9) are flat cases.  
A possible scenario is the picture of the universe beginning with almost completely stiff-fluid domination (fixed point 7 or 8).
In fact, the condition $\xi \rightarrow 0^-$ is required for almost completely dust domination condition (saddle fixed point 5) and kinetic term domination (stiff fluid fixed points 7 and 8). Therefore the requirement is $\xi \rightarrow 0^-$ for it to be physically valid.   
As $\xi\rightarrow 0^-$, the dynamical variable 
$s$ must be negative for an expanding universe, 
The fixed point 8 locates in $s > 0$ region while the fixed point 7 locates in $s < 0$
region. Hence the point 8 is not allowed. In addition, as $\xi \rightarrow 0^-$, it is necessary that $0 < y$ and $A < 0$. The fixed point 7 locates in the allowed region ($s<0, A<0$ and $y>0$) and it is considered as beginning of the evolution.  The universe begins with scalar kinetic term and holographic vacuum energy domination with an almost stiff-fluid EoS. After that it evolves to pass the saddle point 5 with almost-dust dominating EoS. At late time, the evolution goes to the stable point 6 with dark energy dominating EoS.    
The path is therefore $7 \rightarrow 5 \rightarrow 6$.
The fixed points 4 and 9 are not interested as they are not stable for $\xi<0$. 
The stable fixed point 6 corresponds to $  \Omega_{\phi c}+\Omega_{\Lambda c} = 1$. 
as dark energy is contributed from both scalar field and holographic vacuum energy.

Numerical integration can be performed for the autonomous system 
\eqref{eq: autonomous system}. We plot $w_\text{eff}(N)$, $\Omega_\text{m}(N)$, $\Omega_\text{DE}(N)$ and  $\Omega_k(\ln a)$ versus e-folding number, $N=\ln a$. 
 To approach dust or stiff fluid dominations, the NMC coupling  must be small and negative, and the holographic parameter must also be small ($c\rightarrow 0^+$).  For any allowed values of $\xi$ and $c$, $w_\text{eff}$ approaches $-1$ at late times. 
 As NMC coupling becomes more and more negative, dust-dominated EoS gradually disappears. This increases $w_\text{eff}$ values at the fixed points 5 and decreases  $w_\text{eff}$  at the fixed point 7 (FIG. \ref{fig:weff_vary_c}). Increasing of $c$ does not change shape of the $w_{\rm eff}$ but larger $c$ increases $w_\text{eff}$. 
 In FIG. \ref{fig:scalar_density_plot},  the NMC coupling  $\xi$ does not significantly affect $\Omega_\Lambda$, but larger $c$ increases the value of $\Omega_\Lambda$ which  is  nearly constant due to small value of  $\Omega_k$. 
As seen in FIG. \ref{fig:scalar_density_plot}, for larger negative magnitude of the NMC coupling,  $ \Omega_\phi$ is larger in the past.  Larger $c$ reduces the value of $ \Omega_\phi$.
For larger $c$, kinetic part $x$, NMC part $s$ and potential part $y$ are smaller. 
At latest stage of evolution, the potential term finally dominates. 
Very small negative magnitude of $\xi$ and very small $0<c \ll 1$ are preferable in this model. Both scalar field and holographic vacuum energy drives the late-time acceleration. It will be of our interest to investigate further with exponential potential and to perform both kinematical and perturbation analysis against observational data.  

Recent observational analyses \cite{DESI2024,DESI:2024aqx, DESI:2025zgx, Park:2024pew, Notari:2024zmi, Odintsov:2024woi} report some deviations from what is expected in the $\Lambda$CDM model. This findings suggest that dark energy may not be a true cosmological constant but may evolve over cosmic time. Such studies have led to the possibility that a dynamical dark energy model could provide a better fit to current data. 
Our model shows that the non-minimally coupled scalar field and holographic effect can dynamically cause the evolution of $w_{\rm eff}$ with time. In particular, the late-time attractors (fixed point 6), where the scalar field dominates, exhibits behavior consistent with evolving dark energy, depending on the parameter choices. Although we have not fitted our model directly to current observational data in this work, it would be interesting in future studies to compare the model with data to evaluate how well it matches observations.

\section*{Acknowledgments}
Burin Gumjudpai thanks Shinji Tsujikawa and Antonio De Felice for hospitality while visiting Department of Physics of Waseda University and visiting Yukawa Institute for Theoretical Physics of Kyoto University where the work is partially completed.
This research project has been funded by Mahidol University (Fundamental Fund: fiscal year 2025 by National Science Research and Innovation Fund (NSRF)). Amornthep Tita is supported by the Royal Golden Jubilee Ph.D. Programme (RGJ-PhD) scholarship under contact no. PHD/0083/2561.

\end{document}